\documentclass{article}
\usepackage{geometry}
\usepackage{amsmath} 
\usepackage{amssymb} 
\usepackage{amsfonts} 
\usepackage{mathtools} 
\usepackage{hyperref}
\usepackage{caption}
\usepackage{subcaption}
\usepackage{graphicx}
\usepackage{tikz}
\usetikzlibrary{angles,quotes}
\usepackage{float}
\usepackage{environ} 
\usepackage{setspace}

\usepackage[most]{tcolorbox}

\usepackage{biblatex}
\usepackage{color}

\bibliography{references.bib}

\renewcommand{\Re} {\text{Re}}
\newcommand{\Fr} {\text{Fr}}
\newcommand{\ub} {\boldsymbol{u}}

\newcommand{\tb} {\boldsymbol{t}}

\newcommand{\nablab} {\boldsymbol{\nabla}}
\newcommand{\sigmab}{\boldsymbol{\sigma}}
\newcommand{\D}[2]{\dfrac{\text{d} #1}{\text{d} #2}}

\renewcommand{\b}[1]{\boldsymbol{#1}}
\renewcommand{\d}{\text{d}}
\renewcommand{\t}[1]{\text{#1}}
\newcommand\blfootnote[1]{%
    \begingroup
    \renewcommand\thefootnote{}\footnote{#1}%
    \addtocounter{footnote}{-1}%
    \endgroup
}

\title{A Coupled PFEM-DEM Model for Fluid-Granular Flows with Free-Surface Dynamics Applied to Landslides}
\author{Thomas Leyssens, Michel Henry, Jonathan Lambrechts, \\ Vincent Legat, Jean-François Remacle}

\begin{document}
\setstretch{1.125}

\maketitle
\section{Abstract}
Free surface and granular fluid mechanics problems combine the challenges of fluid dynamics with aspects of granular behaviour. 
This type of problem is particularly relevant in contexts such as the flow of sediments in rivers, the movement of granular soils in reservoirs, or the interactions between a fluid and granular materials in industrial processes such as silos. 
The numerical simulation of these phenomena is challenging because the solution depends not only on the multiple phases that strongly interact with each other, but also on the need to describe the geometric evolution of the different interfaces. 
This paper presents an approach to the simulation of fluid-granular phenomena involving strongly deforming free surfaces. The Discrete Element Method (DEM) is combined with the Particle Finite Element Method (PFEM) and the fluid-grain interface is treated by a two-way coupling between the two phases. The fluid-air interface is solved by a free surface model. 
The geometric and topological variations are therefore naturally provided by the full Lagrangian description of all phases.
The approach is validated on benchmark test cases such as two-phase dam failures and then applied to a real landslide problem.
\blfootnote{\textit{Preprint submitted to Journal of Computational Physics \hspace{3cm} December 23rd 2024}}

\section{Introduction}
Fluid mechanics problems involving free surfaces and grains are multifaceted phenomena that require a deep understanding of fluid dynamics and granular interactions. They are particularly relevant to
civil engineering, earth science, and many industrial applications.

\begin{figure}[h!]
    \begin{center}
        \includegraphics[width=.7\textwidth]{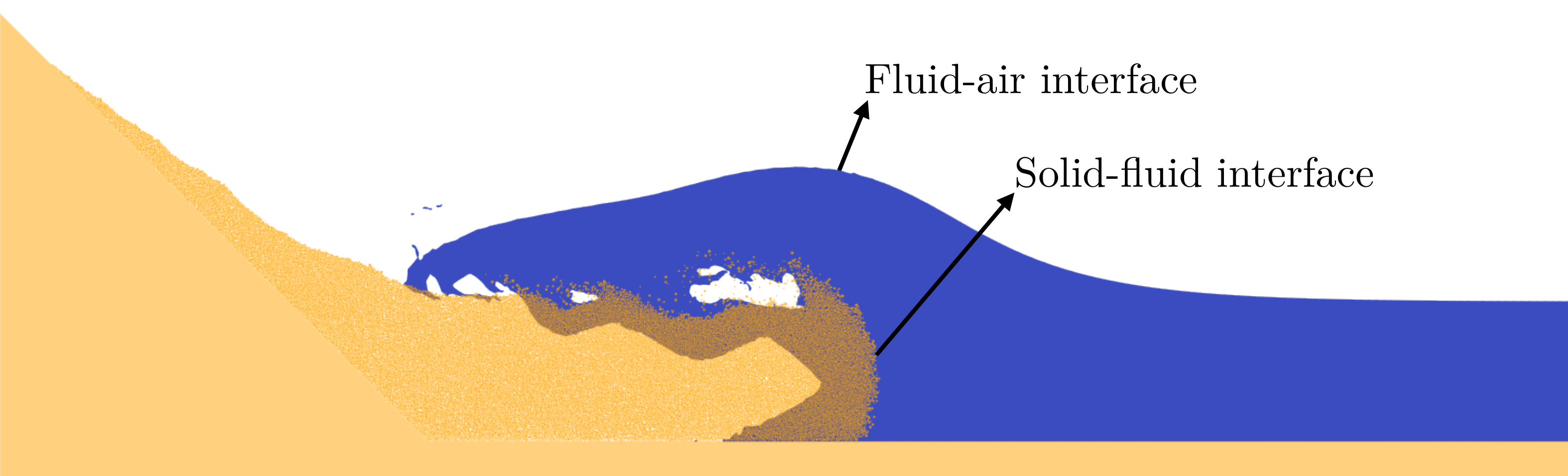}
        \caption{In this study, two types of interfaces are a source of complexity: the free-surface between fluid and air, and the solid-fluid interface.}\label{fig:interfaces}
    \end{center}
\end{figure}

For example, let us consider landslides. 
These natural phenomena often have catastrophic consequences for people, and understanding the dynamics and predicting the potential impact in advance can help prevent the worst. 
Landslides occur when soils on steep slopes become unstable. 
In some cases, these steep hillsides are located above water basins, resulting in a hard impact of the soil on the water, causing significant impulse waves on the water.
In other cases, known as submarine landslides, the soil detaches below the water level, creating waves on the water surface.
Both cases can be referred to as landslide tsunamis \cite{ward}, and it remains a difficult challenge to simulate the creation of these waves near the source, in order to predict their propagation\cite{abadieVOF}.
In these phenomena, the \emph{two main interfaces} are solid-fluid between the moving soil and water, and fluid-air at the surface of the water.
Figure \ref{fig:interfaces} is an illustration of the Lituya Bay landslide and tsunami, a case that will be studied in this work.
This illustrates the complexity of the two types of interfaces.

The numerical simulation of the evolution of these multiple strongly interacting interfaces is difficult because it is non-linear. The interaction laws between the fluid and granular phases are non-linear, but this is not the only non-linearity. Another important non-linearity is related to the geometry of the interfaces: we want to model moving interfaces with possible changes in topology, including nucleation and annihilation.

Our problem thus has two types of interfaces corresponding to specific physical couplings. We will now describe the state of the art in modeling these couplings individually and finally propose a new method that has all the right qualities to simulate all the couplings at once. 

\subsection{Fluid-air coupling -- two phase flows}
For the fluid-air coupling, several approaches have been proposed in the literature.
Multi-fluid formulations such as volume-of-fluid \cite{hirtVOF} or level-set \cite{sussmanlevelset} approaches simulate both phases completely. 
However, the cost of doing so can quickly become prohibitive. 
In fact, the time scales of the dynamics in the lighter, less viscous fluid are often much shorter than in the heavier, more viscous fluid. 
Much unnecessary work would be spent solving the air phase, which has no significant effect on the flow.
In addition, many formulations consider smooth interfaces between the phases because it is difficult to follow topological changes in a sharp manner. 
This can lead to a loss of accuracy in the representation of the two phases as a result of diffusive effects.

Another strategy consists in simulating only the heavier fluid. 
This is an attractive solution, as the equations only require to be solved on one part of the domain, and the time step can be chosen at the scale of the dynamics of the heavier fluid.
The fluid-air interface thereby becomes a \emph{free-surface}, on which the kinematics are dictated by the fact that there is no flow through the interface.
The main challenge becomes geometrical.
Indeed, the motion of the interface must be tracked and topological changes such as two parts of the domain colliding or separating, should also be accurately detected.

Here, Lagrangian approaches are more common, as the degrees of freedom move with the flow.
Lagrangian methods can be separated into mesh-based and meshless methods.
Meshless methods, typically considered as particle methods, do not require the generation of a mesh, and field gradients are computed using neighborhood samples around the particles.
The smoothed particle hydrodynamics (SPH)\cite{sph}, the material point method (MPM)\cite{sulsky}, and the lattice Boltzmann method (LBM)\cite{pasquaLBM} are popular examples of these methods. 
Mesh-based methods, on the other hand, make use of a mesh connecting the particles to solve the equations of motion. 
These methods benefit from the strong mathematical background of mesh-based methods, such as the finite element method, to solve the equations and apply boundary conditions. 
Arbitrary Lagrangian-Eulerian (ALE)\cite{ale} belong to this category. 
They are a good method of choice for simulating deforming domains thanks to great conservation properties, but become complex and prohibitive when high levels of topological changes are desired.
Another challenge in these methods remains to efficiently generate meshes onto these particles. 
With recent advances in mesh generation techniques\cite{celestin}, however, this step should no longer be considered a bottleneck to the entire simulation process.

In this work, the Lagrangian mesh-based method known as the particle finite element method, or PFEM\cite{pfem}, is used. 
In brief, the PFEM solves the equations of motion using a finite element formulation, and then displaces the nodes of the mesh, the particles, based on the obtained displacement or velocity field. 
As this Lagrangian displacement may lead to great distortions of the mesh, especially in highly deforming domains such as fluids, the mesh is discarded at each time step, keeping only the nodes, on which a new triangulation is performed.
The free-surface boundary is then tracked through the use of an indicator function that defines the shape of the domain.

\subsection{Fluid-solid coupling}
The second interface, between the solid and the fluid, arises as the solid phase is driven into the flow.
The soil can either be modeled as a continuum or in a discrete manner.
In the continuum approach, the mixture composed of fluid and soil is modeled as a non-Newtonian fluid with proper parametrization of mixture stress tensor.
This approach is commonly referred to as a two-fluid model \cite{baumgartenGeneralFluidSediment2019} which provides a good approximation of macroscopic behavior of the soil.
On the other hand, the soil can be modeled as a discrete phase, composed of individual grains.
This approach is often referred to as the discrete element method (DEM) which solves the equations of motion for each individual grain.
While grains are moving in the flow, the fluid exerts a force on each grain which alters its dynamics and collisions may occur between grains or the boundaries.
To resolve a contact impulse, two methods are commonly used: the smooth contact dynamics and the non-smooth contact dynamics, also called the hard sphere model.
In the smooth contact dynamics, the contact forces are computed using an explicit penalty method based on the overlap of the grains \cite{kruggel-emdenReviewExtensionNormal2007}.
In the non-smooth contact dynamics, contact impulses are iteratively obtained by prohibiting grain overlaps \cite{jeanNonsmoothContactDynamics2001}.
Both approaches are able to efficiently and accurately describe contact forces while integrating frictional and cohesive forces if needed.
Recalling Newton's second law, as the fluid exerts a force on the grains, the grains exert the same force on the fluid.
To take into account the grain presence in the fluid, volume-averaged methods can be used.
One of the most popular ones is the CFD-DEM method, which solves the volume-averaged Navier-Stokes equations (VANS) on a fixed mesh and computes the forces exerted by the grains on the fluid using drag correlations \cite{geitaniHighorderStabilizedSolver2023a}.
Using a fixed mesh can be quite restrictive in some cases.
Indeed, the applications of interest here consider a Lagrangian representation of the fluid, and the model should therefore be capable of solving the coupled problem on non-constant meshes.
Moreover, the averaging process assumes the grains to be smaller than the mesh size. 
In this work, this limitation is overcome by computing the exact overlap between grains and the mesh, resulting in an accurate and conservative void fraction \cite{clarkeInvestigationVoidFraction2018, stroblExactCalculationOverlap2016}. 

\subsection{Fluid-Granular Flows with Free-Surface Dynamics}
The present work aims at simulating granular free-surface flows. 
The applications require to have sharp interfaces between the fluid, the air and the solid, and allow large deformations of the domain. 
Different approaches for simulating free surface flows coupled to a solid granular formulation have been proposed.
Eulerian-Lagrangian methods such as FEM-DEM couplings with a multi-phase flow formulation have been proposed, for instance \cite{mao}. 
In their approach, the free-surface interface is captured using a conservative levelset method, while the solid-fluid interface is managed by an immersed boundary method. 
Such approaches benefit from high levels of accuracy and good conservation properties. 
The fact of requiring a levelset function to represent the fluid domain, however, may increase the computational cost and strong topological changes are often complicated to capture.

Lagrangian-Lagrangian formulations have also been proposed, such as a moving particle semi-implicit (MPS) formulation for the fluid, coupled with the DEM method for the solid granular phase\cite{xie}.
Similarly, an SPH-DEM coupled formulation was proposed by \cite{sun}. 
In both methods, local gradients are defined through the use of a smoothing kernel based on neighboring particles, and therefore do not require the proper definition of a mesh. 
An important aspect of the coupling between the fluid and solid phases in this method is that a large number of fluid particles is required around each solid particle, and this ratio should be kept constant. 
Simulating solid impacts on the free-surface (\textit{i.e.}, solid particles entering the fluid) may therefore be challenging. 

In \cite{vajont}, a fully Lagrangian approach was proposed to simulate landslide phenomena, in which both phases are considered as a continuum. 
More specifically, both phases are simulated with the particle finite element method, but two different constitutive models are considered for each phase. 
While the water phase is modeled as a standard Newtonian fluid, the landslide phase is parametrized using a frictional viscoplastic model.
Three-dimensional results were demonstrated and used to simulate the 1963 Vajont rockslide and tsunami.

Finally, a PFEM-DEM formulation has been proposed by Franci et al. \cite{franci} for a one-way coupling for particle-laden flows. 
Since the particles are sufficiently small and disperse such as to consider their effect on the fluid as negligible.
Hence, the approach of Franci et al. is very similar to the method proposed in this paper, albeit without the two-way coupling.

\begin{figure}[H]
    \begin{center}
        \includegraphics[width=\textwidth]{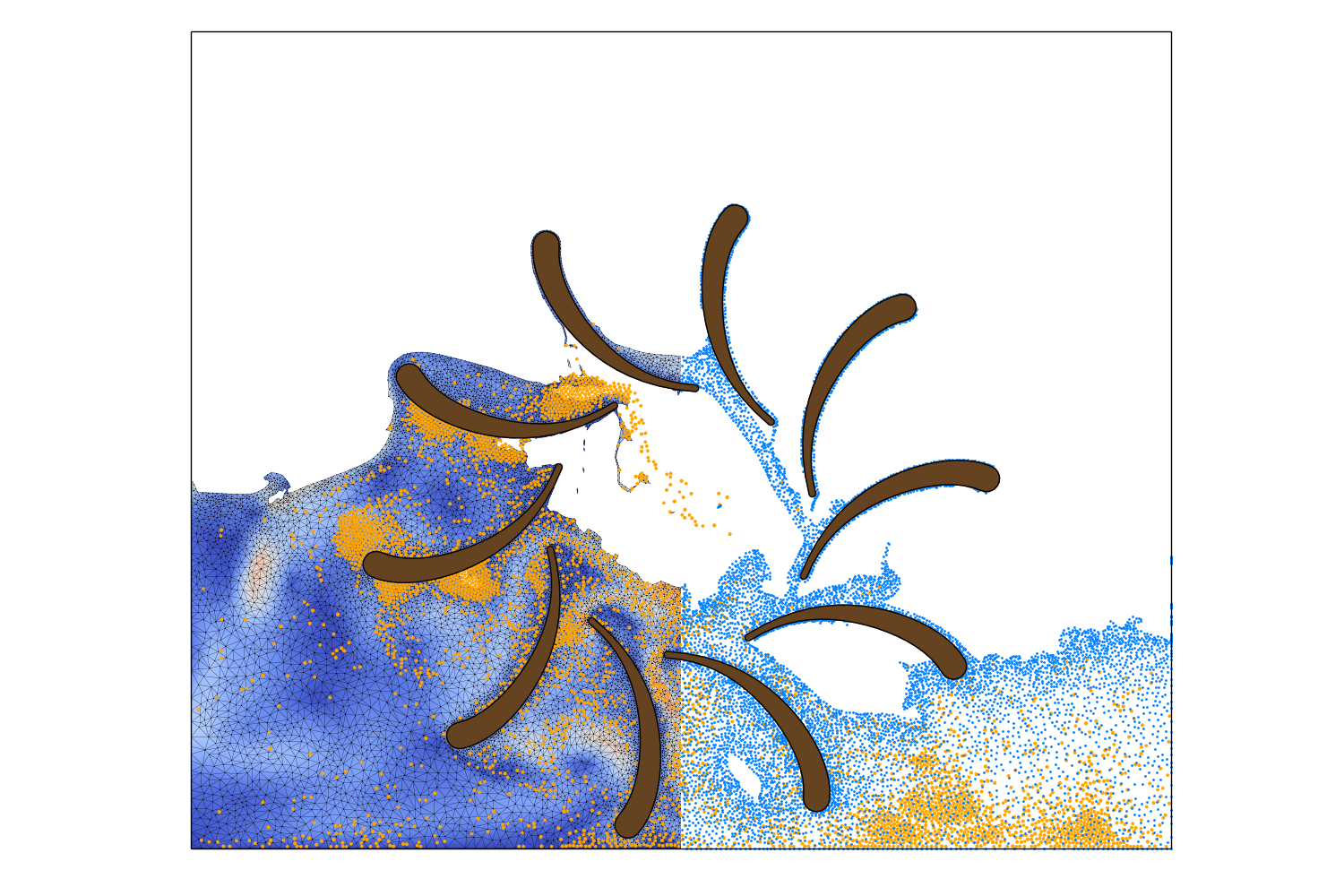}
        \caption{We present a fully coupled PFEM-DEM formulation for the simulation of fluid-grain flows with strong domain deformations.}\label{fig:wheel}
    \end{center}
\end{figure}

In this paper, we present a new PFEM-DEM formulation to simulate two-way coupled fluid-grain problems with large deformations of the fluid domain. 
Figure \ref{fig:wheel} demonstrates the potential of our solver. 
Deforming free surfaces, fluid-grain coupling, complex evolutions of the domain and moving boundaries are all within the capabilities.
This method is of great interest for simulations such as landslide-generated impulse waves.
Although the rotating turbine is an illustrative example, the present paper aims to validate our approach to landslide-related problems. 
Indeed, thanks to the Lagrangian formulation of the fluid, free-surfaces are inherently captured, allowing to simulate strongly dynamic phenomena such as impulse waves. 
The discrete element method, used for the granular phase, is adequate for the landslide phenomena simulated in this paper, but also has the capability to simulate phenomena at much smaller scales, understating the versatility of the approach.

We first present the fully coupled formulation, followed by two benchmarks to validate the approach. 
Finally, we apply the method to the 1958 Lituya Bay landslide and tsunami, a real-life application.

\section{The Physical Problem}
In this section, the governing equations of the physical problem are presented. 
The fluid flow within the multi-phase system is governed by the Volume-Averaged Navier-Stokes (VANS) equations. 
These fluid conservation equations are derived by averaging the Navier-Stokes equations over a control volume containing both phases, similar to a two-fluid approach. 
The Lagrangian formulation of the fluid phase is used to track the fluid material points. 
This description provides a natural way to capture the domain deformation and the free surface dynamics.
The motion of individual grains is modelled using the Newton-Euler equations, assuming the grains to be rigid bodies with a cylindrical shape. 
Interactions between grains, as well as between grains and the boundaries, are handled by the Non-Smooth Contact Dynamics (NSCD) method \cite{jeanNonsmoothContactDynamics2001}. 
To represent the grains in the weak form of the VANS equations, a least-squares projection is used.
The coupled problem is thus described by the following set of equations:

\vspace{0.5cm}
\noindent
\begin{center}
\begin{minipage}[t]{0.6\textwidth}
    \begin{tcolorbox}[enhanced,
        attach boxed title to top center={yshift=-3mm,yshifttext=-1mm},
        coltext=black,
        colback=white,
        colframe=black,
        colbacktitle=cyan,
        title=\color{black} Fluid,
        left*=-5pt,
        right*=5pt,
        height=5cm
        ]
        \begin{align}
            &\D{\b{x}}{t} = \b{u}, \\[0.65cm]
            &\displaystyle\int_{} \ \epsilon \rho \D{\ub}{t} \cdot \hat{\b{u}}\ \d \b{x} = \displaystyle\int_{} \bigg( \nablab \cdot \sigmab  - (1-\epsilon)\boldsymbol{f}_{} + \epsilon \rho \boldsymbol{g} \bigg)    \cdot \hat{\b{u}}\ \d \b{x},\label{eq:vans_momentum}\\[0.65cm]
            &\displaystyle\int_{} \ \bigg(\nablab \cdot \big(\epsilon \ub + (1 - \epsilon) \b{v}\big)\bigg) \hat{p}\ \d \b{x} = 0,\label{eq:vans_mass}
        \end{align}
    \end{tcolorbox}
\end{minipage}
\hfill
\begin{minipage}[t]{0.39\textwidth}
    \begin{tcolorbox}[enhanced,
        attach boxed title to top center={yshift=-3mm,yshifttext=-1mm},
        coltext=black,
        colback=white,
        colframe=black,
        colbacktitle=orange,
        title=\color{black} Grains,
        left*=-5pt,
        right*=5pt,
        height=5cm
        ]
        \begin{align}
            &\D{\b{X}_{i}}{t} = \b{V}_{i}, \label{eq:grains_displacement}\\[0.3cm]
            &m_{i} \D{\b{V}_{i}}{t} = \boldsymbol{F}_{i}^\t{c} + \boldsymbol{F}_{i} + m_{i} \boldsymbol{g},\label{eq:grains_translation} \\[0.2cm]
            &\boldsymbol{I}_i \D{\boldsymbol{\Omega}_{i}}{t} = \boldsymbol{M}_{i}^\t{c} + \b{M}_{i},\label{eq:grains_rotation}
            \\[0.3cm]
            &\mathcal{C}(\b{F}^\t{c}, \b{M}^\t{c}, \b{V}, \b{X}) = 0,
        \end{align}
    \end{tcolorbox}
\end{minipage}

\begin{minipage}[t]{0.65\textwidth}
\begin{tcolorbox}[enhanced,
    attach boxed title to top center={yshift=-3mm,yshifttext=-1mm},
    coltext=black,
    colback=white,
    colframe=black,
    colbacktitle=gray!50!white,
    title=\color{black} Coupling,
    height=4.5cm,
    ]
    \begin{align}
        &\displaystyle\int_{}\ \left(1 - \epsilon\right)\ \hat{p}\ \d \b{x}  = \displaystyle\int_{}\ \displaystyle\sum_{i} H_i \ \hat{p}\ \d \b{x},\label{eq:void_fraction}\\
        &\displaystyle\int_{}\ \left(1-\epsilon\right) \b{f}_{} \cdot \hat{\ub}\ \d \b{x}  = \displaystyle\int_{}\ \displaystyle\sum_{i} \frac{H_i \boldsymbol{F}_{i}}{V_i}  \cdot \hat{\ub}\ \d \b{x},\label{eq:fluid_grain_force}\\
        &\displaystyle\int_{}\ \left(1-\epsilon\right )\b{v} \cdot \hat{\ub}\ \d \b{x}  = \displaystyle\int_{}\ \displaystyle\sum_{i} H_i \b{V}_{i} \cdot \hat{\ub}\ \d \b{x}.\label{eq:fluid_grains_velocity}
    \end{align}
\end{tcolorbox}
\end{minipage}
\end{center}

\vspace{0.5cm}

\noindent
In these equations, $\b{x}$ is the fluid material point position, $\b{u}$ is the fluid velocity, $p$ is the pressure, $\rho$ is the fluid density, $\eta$ is the fluid dynamic viscosity, $\epsilon$ is the void fraction, $\sigmab$ is the fluid stress tensor, $\boldsymbol{f}$ is the fluid-grain interaction force, $\boldsymbol{g}$ is the gravitational acceleration, $\b{X}$ is the grain position, $\b{V}$ is grain velocity, $m$ is the grain mass, $\boldsymbol{F}^\t{c}$ is the contact force, $\boldsymbol{F}$ is the fluid-grain interaction force, $\boldsymbol{I}$ is the inertia tensor, $\boldsymbol{\Omega}$ is the angular velocity, $\boldsymbol{M}^\t{c}$ is the contact torque, $\boldsymbol{M}$ is the fluid-grain interaction torque, $\mathcal{C}$ is the global contact dynamics operator.
Finally, $H_i$ is the $i$-th grain's Heaviside function, defined to be equal to 1 inside grain $i$'s volume $V_i$, and 0 elsewhere.

The domain $\Omega$ is tracked by the fluid material points $\b{x}$.
The material derivative is denoted by $\frac{\text{d} \cdot}{\text{d}t}$ which does not produce any convective term in the momentum equation as it is treated by the nodal displacement.
The fluid stress tensor $\sigmab$ is modelled as:
\begin{equation}
    \sigmab = -p \boldsymbol{\delta} + \epsilon \eta \left( \nablab \ub + \nablab^T \ub   \right),
\end{equation}
where $p$ is the pressure, $\eta$ is the dynamic viscosity, and $\boldsymbol{\delta}$ is the identity tensor.
The fluid velocity and pressure are solutions of the weak form of the VANS equations, Equations \eqref{eq:vans_momentum} - \eqref{eq:vans_mass}, where $\hat{\ub}$ and $\hat{p}$ are the test functions associated with the velocity and pressure fields, respectively.

The grains are modelled as rigid bodies.
The position of each grain $i$ is tracked by its centre of mass, Equation \eqref{eq:grains_displacement}.
The translational and rotational velocities are solutions of the Newton-Euler equations, Equations \eqref{eq:grains_translation}-\eqref{eq:grains_rotation}.
Contacts between grains and the boundaries are modelled using the Non-Smooth Contact Dynamics (NSCD) assuming inelastic collisions. 
The abstract contact dynamics operator $\mathcal{C}$ is used to enforce the contact conditions.
For each contact, these contact constraints are: the Signorini condition, Coulomb's friction law and the momentum balance at the contact point. 
The Signorini condition ensures that the grains do not overlap at the end of the collision.
Coulomb's friction law ensures that, at the contact point, the relative velocity of the grains remains within the tangential cone. 
The contact dynamics describes the relative velocity of two grains after the collision to the contact impulse.
A detailed description of this methodology can be found in \cite{duboisContactDynamicsMethod2018,coppinNumericalAnalysisDrag2021,constantImplementationUnresolvedStabilised2018}.

\begin{figure}[H]
    \centering
    \includegraphics[width=0.6\textwidth]{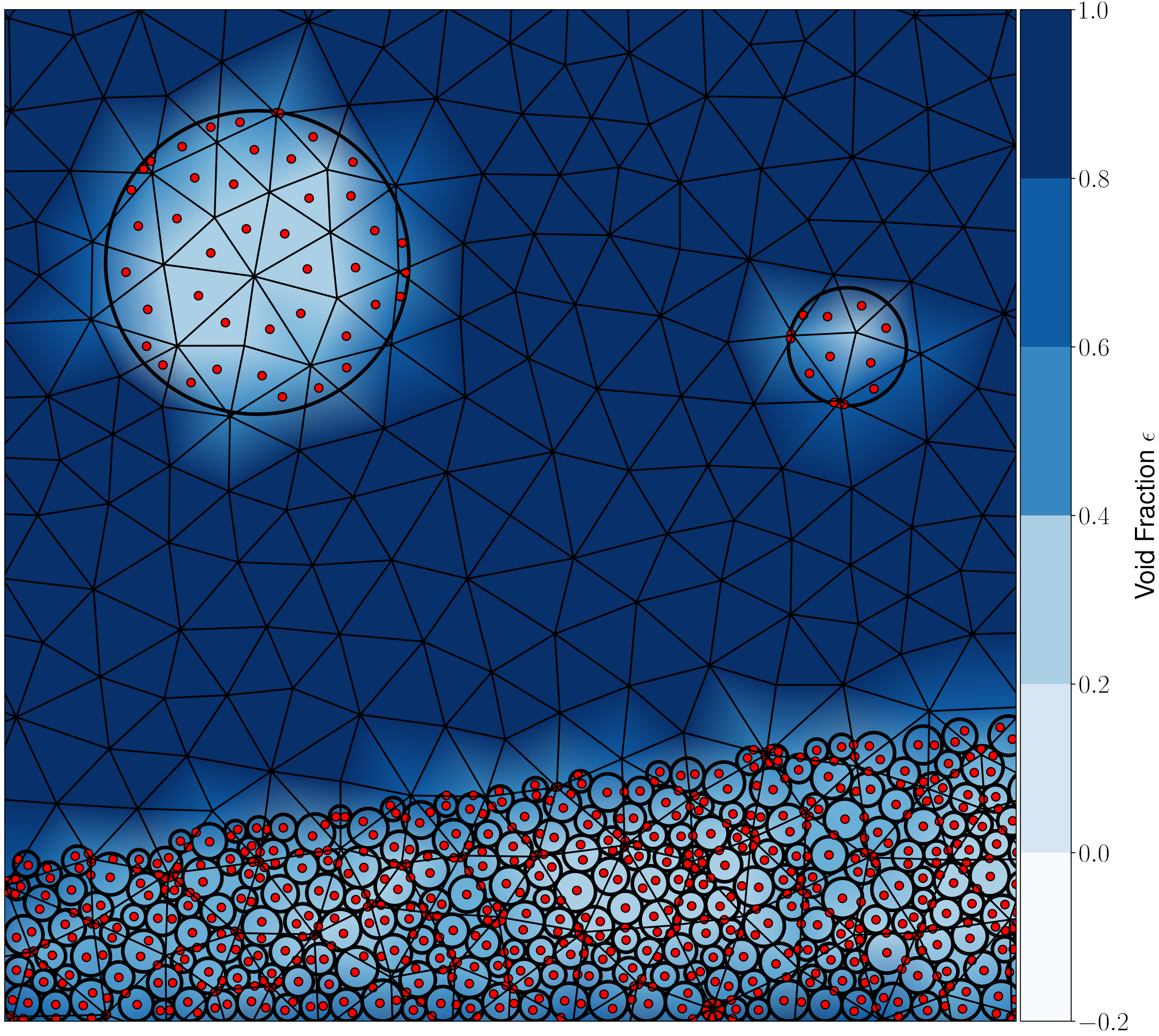}
    \caption{Illustration of the fluid-grains coupling. 
    Grains are represented as a continuous field using an $L_2$ projection.
    The integral over the disc is approximated by the sum of the integrals over the elements containing the grain. 
    The red dot represents the centroid of each overlapped sub-element.
    The void fraction is indicated by the colour map.}\label{fig:fluid_grains_coupling}
\end{figure}

Since the grains are defined in a purely discrete manner, a projection step is required to include them in the weak form of the VANS equations. 
This is achieved through an $L_2$-projection of the grains over the fluid domain, Equations \eqref{eq:void_fraction}-\eqref{eq:fluid_grains_velocity}.
The fields of interest are the void fraction, \textit{i.e.} the fluid volume fraction, the fluid-grain interaction force, and the grain velocity.
Figure \ref{fig:fluid_grains_coupling} illustrates the projection of grain volume to the fluid domain in order to compute the void fraction.
A grain is divided into sub-elements, and the integral over the grain is approximated by the sum of the integrals over the sub-elements.
Details on the spatial representation of the grains are discussed in the next section.

To model the fluid-grain interaction occuring at the grain scale, constitutive laws must be introduced.
The fluid-grain interaction force is modelled as the sum of the drag force and the macroscopic stress tensor acting on grain $i$,

\begin{equation}
    \boldsymbol{F}_{i} = V_i \nabla \cdot \boldsymbol{\sigma} + \boldsymbol{F}_{\t{d}}, \label{eq:solid_fluid_force}
\end{equation}
with $\boldsymbol{F}_\t{d}$ the drag force. 
The only force considered in this work is the drag force acting on the grain. Other forces such as the lift force or the Basset history force are neglected.
Therefore, no torque is applied to the grains by the fluid phase, $\b{M}_i \approx 0$.
To estimate the drag force acting on the grain, the Dallavalle correlation and the Di Felice voidage function are used \cite{dallavalleMicromeriticsTechnologyParticles1943} \cite{difeliceVoidageFunctionFluidparticle1994},

\begin{equation}
    \boldsymbol{F}_\t{d} \approx \underbrace{\epsilon^{-2.8} {\left(0.63\sqrt{\epsilon\Re} + 4.8\right)}^2 \dfrac{\pi d}{4} \eta}_{\triangleq \gamma} \left(\b{u} - \b{V}_i\right),\label{eq:drag_correlation}
\end{equation}
where $d$ is the grain diameter and $\Re$ the Reynolds number,

\begin{equation}
    \Re = \dfrac{\rho d \lVert \b{u} - \b{V}_i \rVert}{\eta}.
\end{equation}
No angular velocity is considered to compute the drag force.

\section{A fully Lagrangian scheme}
\begin{figure}[ht!]
    \centering
    \includegraphics[width=\textwidth]{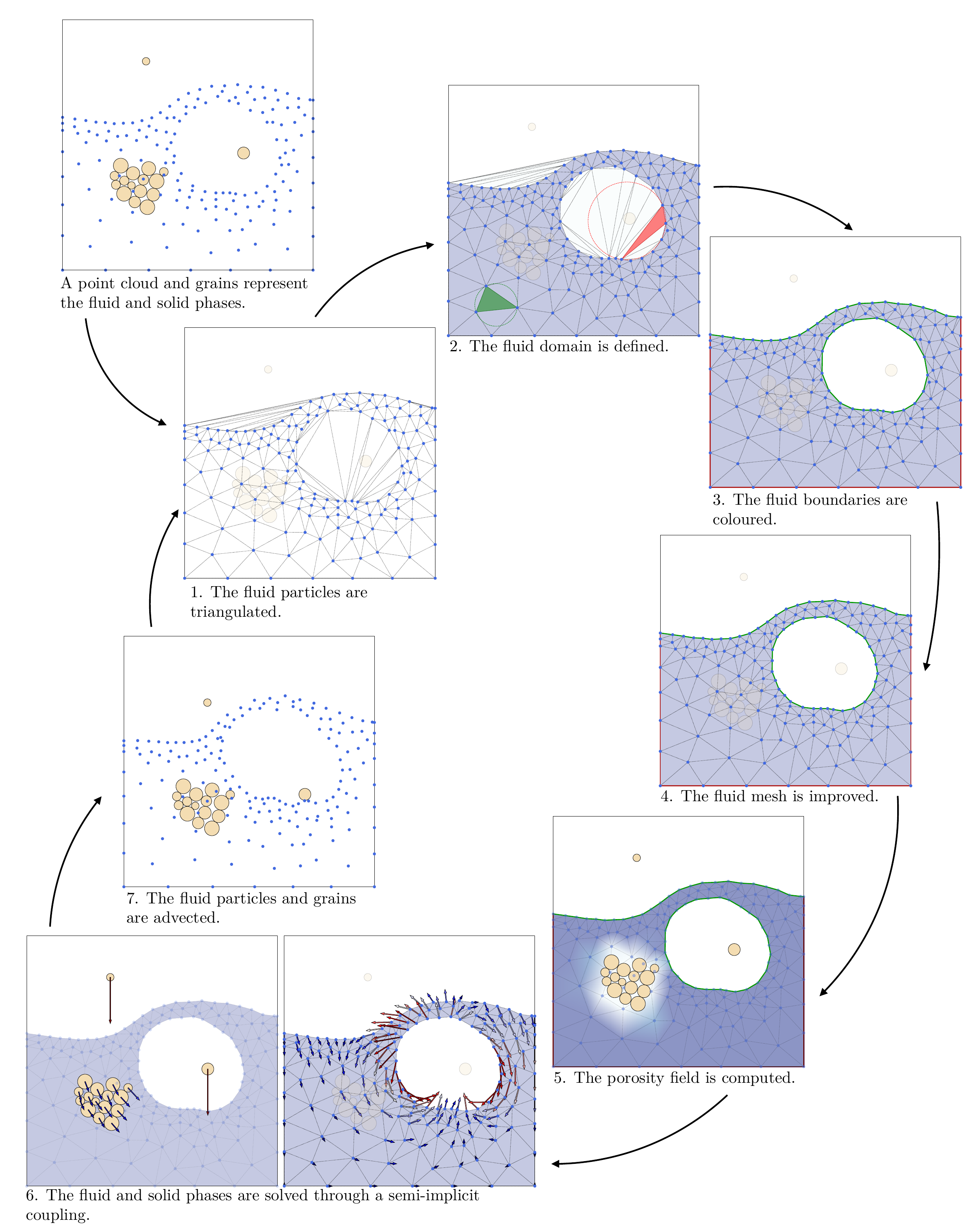}
    \caption{A schematic of the algorithm.}\label{fig:scheme}
\end{figure}

The simulations of interest in the current study occur in extremely strongly deforming domains. 
This is the case for both the solid phase and the fluid phase. 
For this reason, considering Lagrangian formulations, inherently capable of describing the motion and deformations of the domains, is crucial. 
The decision was therefore made to adopt a particle-based formulation for both phases.

This not only allows to track topological changes in both phases, but also to naturally couple them together.
Hence, both phases are initially represented by two separate sets of particles: the grains and the fluid.
This is the starting point of the overall algorithm of the coupled method, presented in Figure \ref{fig:scheme}.
This figure will help the reader understand the algorithm. 
In what follows, the fluid discretization approach is presented first, followed by the solid-fluid coupling procedure to integrate the equations in time and space.
Finally, the advection steps for both the fluid particles and solid grains are described. 
\subsection{Fluid particles triangulation}

Capturing highly deforming fluid domains requires a specific technique that inherently follows the evolution of the geometry.
The particle finite element method has been specifically designed for such flows. 
In this Lagrangian method, all the relevant information is passed on from one time step to another through the particles. 
Then, with the aim to solve the equations of motion, a Delaunay triangulation of these particles is performed, resulting in a mesh of the convex hull (Figure \ref{fig:scheme}: point 1).

\subsection{Domain definition}
\begin{figure}[H]
    \centering
    \includegraphics[width=\textwidth]{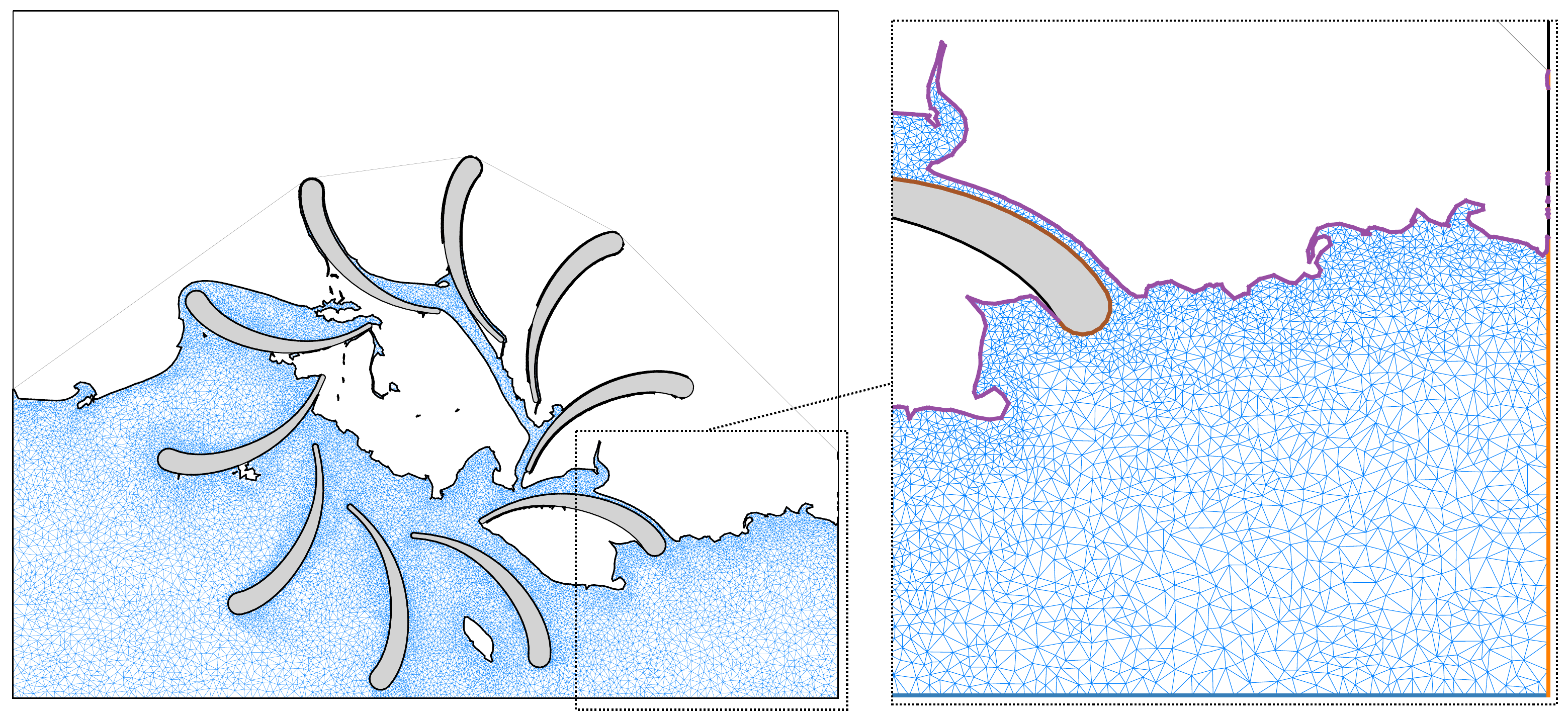}
    \caption{Chaotic free surface flows require the method to be able to track complex topological changes and to detect boundaries.}\label{fig:mill}
\end{figure}

The particles are displaced after each time step in such a way that it is impossible to simply deform the initial mesh.
In addition to the \emph{tracking} of the interfaces in a continuous manner, the topological changes appearing near free surfaces must also be taken into account.
We thus only rely on a point cloud $S$ to represent the fluid domain and its boundary. 
It is possible (and fast) to triangulate $S$ and the boundary of this triangulation will give the new ‘shape’ of the fluid domain. The \emph{convex hull} $H(S)$ of a point cloud $S=\{s_1,\dots,s_n\}$ is a clear and perfectly defined concept. However using $H(S)$ as the \emph{oracle} that determines the boundary of the fluid domain is not very interesting. Being always convex, it is obviously very restrictive. Figure \ref{fig:mill} illustrates the importance of having a method that accurately covers something else than a convex hull. 

The very notion of the boundary of a point cloud is not unique. There are a large number of methods that can be used to calculate the boundary of $S$, one being called the $\alpha$-shape of $S$ \cite{edelsbrunner}. The concept of $\alpha$-shape is closely linked to the Delaunay triangulation DT$(S)$. 
The $\alpha$-complex of $S$ is the set of triangles of DT$(S)$ whose circumradii are smaller than a given length. 
The $\alpha$-shape of $S$ is the boundary of the $\alpha$-complex.

To capture the evolution of the shape of the fluid domain, we compute DT$(S)$ and subsequently remove
triangles that are too large (see Figure \ref{fig:scheme}: point 2).
We verify that the circumradius $R_e$ of 
each element satisfies the following criterion: 
\begin{equation}
    R_e < \alpha \cdot h(\mathbf x). \label{eq:alpha-shape}
\end{equation}
Here, $h(\mathbf x)$ is a mesh size field that defines
how acceptable circumradii are distributed over the domain.
In this context, the parameter $\alpha \geq 1$ 
corresponds to a tolerance. 
Size fields are continuous functions of the position $\bf x$ and meshes are discrete, so we allow some long edges/large elements to be acceptable.

\subsection{Boundary colouring}

In Lagrangian formulations, it is not possible to define the boundaries only once at the beginning of the simulation.
Indeed, since the particles follow the deformation of the fluid and its topological changes, interactions with the walls surrounding the fluid will also change over time. 
This is even more critical in cases of moving boundaries, as illustrated in Figure \ref{fig:mill}. 
From the $\alpha$-shape algorithm, the boundaries of the fluid domain are automatically obtained. 
However, the distinction between different boundaries such as walls, moving bodies and free-surfaces is not automatic, and therefore, the fluid boundary edges must be coloured according to which boundary wall they belong.
In Figure \ref{fig:scheme}, point 3, fluid boundaries belonging to the wall are coloured in red, and free-surfaces in green. 

In the method proposed here, walls are not explicitly meshed, and the fluid only recognises them geometrically. 
It is therefore necessary to be able to detect when fluid particles are situated on these boundaries, and to forbid them of crossing walls.
To achieve this, a representation of the boundaries is stored in an octree \cite{octree} data structure. 
This allows to make quick searches and to easily compute distances between particles and walls.
Considering two particles that form a boundary edge, if these two particles are within a given tolerance of a solid wall, this boundary edge is coloured as a wall.
If not, this boundary edge is considered a free-surface edge.
The zoom of Figure \ref{fig:mill} illustrates the colouring of the boundary edges.

Applying boundary conditions then becomes straightforward, and can be done in the same manner as any classical finite elements formulation. 
The issue of fluid particles crossing solid boundaries is addressed in the advection step, described in section \ref{advection}.

\subsection{Mesh adaptation}

Mesh adaptation is crucial in the PFEM. 
Indeed, since the definition of the fluid domain relies on the $\alpha$-shape, one must ensure that the elements inside the fluid domain respect the given size field and respect condition \eqref{eq:alpha-shape}.
Due to the Lagrangian motion of the particles, the triangulation will be deformed.
This can lead to very poor quality elements in the mesh, and consequently a poor definition of the fluid domain and its boundaries.

In \cite{leyssens}, an adaptive mesh refinement technique has been proposed for the PFEM to address this challenge. 
The size field $h(\mathbf x)$, used to filter elements in the $\alpha$-shape procedure, is chosen to be highly refined near the free surface, and coarser at a distance from these boundaries.
Indeed, since it is near the free surface that the topological changes occur, it is desirable to have greater accuracy in these regions. 
After defining the boundaries of the fluid domain through the $\alpha$-shape, an algorithm broadly inspired by the constrained Delaunay technique of Chew \cite{chew} is then used to meet the quality and size criterion imposed by the size field. 
This method allows to maintain high quality elements at all times, and ensures a smooth and continuous representation of the free surface.
An interpolation of the solution is performed on newly inserted nodes. 
\subsection{Spatial representation of grains in a fluid}
    Since grains are represented as discrete entities, they are projected onto the continuous representation of the fluid.
    Grains are projected through an $L_2$ projection with the same basis functions as those used in the weak form of the fluid equations, defined by Equations \eqref{eq:void_fraction}-\eqref{eq:fluid_grains_velocity}.

    To compute the void fraction and the averaged velocity of the grains, the mass-matrix, which arises on the left hand-side of the weak form, is mass-lumped \cite{chen1985lumped, geitaniHighorderStabilizedSolver2023a}.
    Conversely, the fluid-grain interaction force is computed locally for each grain and then applied to the fluid.
    This is done to ensure Newton's third law.

\begin{figure}[H]
        \centering
        \includegraphics[width=0.5\textwidth]{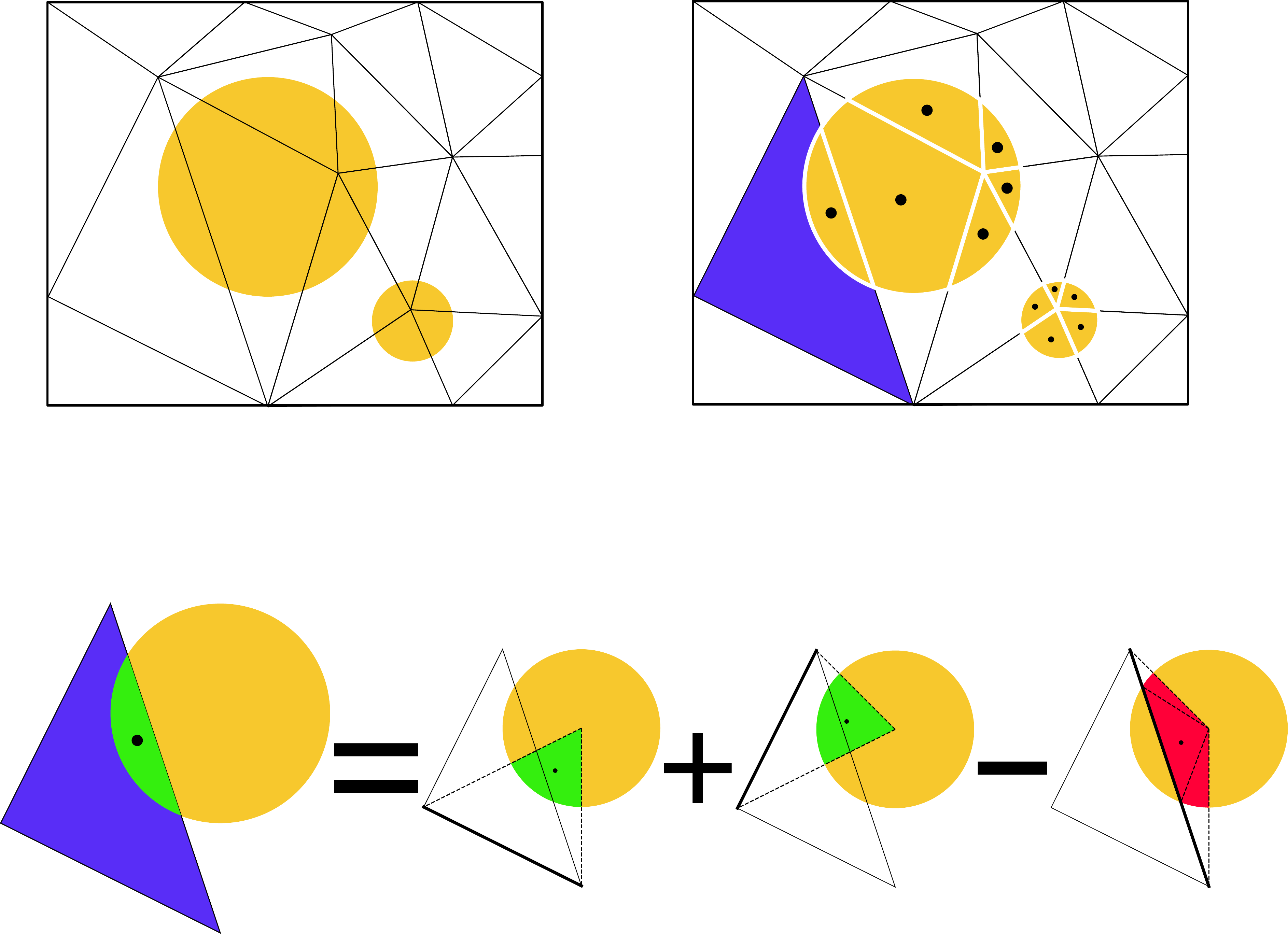}
        \caption{The Heaviside function of a disc is integrated over the fluid mesh. 
        Subdomains are defined by the facets of each mesh elements.
        The centroid of each subdomain is computed by iterating over the element edges.
        }\label{fig:overlap_edges}
    \end{figure}

    To represent a grain, the disc is subdivided by mesh facets, generating sub-elements.
    The centre of mass of each sub-element is computed analytically by iterating over the edges of the overlapped element. 
    For each edge, a triangle is constructed using the edge and the grain's centre of mass. 
    The exact intersection between this triangle and the disc is computed, along with the centroid of the intersection as it is only composed of triangles or disc sections.
    Using these centroids, the centre of mass of each sub-element is determined, on which the integrand is then evaluated.
    Figure \ref{fig:overlap_edges} illustrates this procedure.

    This approach allows for the exact integration of a linear integrand over the intersection, ensuring that the void fraction is computed precisely while maintaining positive values bounded by unity. 
    Additionally, if a grain spans multiple elements, the interaction force is distributed across the elements containing the grain, rather than being concentrated at the grain's centre of mass. 
    This distribution smooths out jumps in the fluid-grain interaction force when a grain enters or leaves an element.

\subsection{A semi-implicit temporal scheme}
    Since the fluid-grain interaction force is a source term in the momentum equation, it is crucial to ensure the stability of the numerical scheme as its magnitude can be significant.
    This is achieved by using a semi-implicit scheme to estimate the fluid-grain interaction force.
    The grain index $i$ is dropped for clarity.
    The force is linearised and a prediction of the grain velocity at the integration point is made based on the grains dynamics, Equation \eqref{eq:grains_translation},

    \begin{equation}
        m\dfrac{\b{V}^* - \b{V}^{n}}{\Delta t} = {\b{F}^{\t{c}}}^{n}\underbrace{-V\b{\nabla} p^{n+1} + \gamma^n \left(\b{u}^{n+1} - \b{V}^*\right)}_{\b{F}^*} + m \b{g}  \\
    \end{equation}
    where subscript `$n$' denotes the value at the previous time step, subscript `$n+1$' the value at the current time step, and subscript `$*$' the predicted value.
    As the contacts have not been resolved yet, the contact force is taken as the one estimated at the previous time step.
    It leads to the following prediction of the grain velocity at the integration point,

    \begin{equation}
        \b{V}^* = \b{V}^n + \dfrac{\Delta t}{m + \gamma^n\Delta t}\left({\b{F}^\t{c}}^n - V \b{\nabla} p^{n+1} + \gamma^n \left( \b{u}^{n+1} - \b{V}^n \right) + m \b{g} \right).
    \end{equation}
    This estimate is then used to compute the drag force acting on the fluid and to project the grain mass flux in the continuity equation, Equation \eqref{eq:vans_mass}.
    An implicit Euler scheme is used to integrate stress tensor leading to the following time discretisation of the VANS equations,

    \begin{align}
        &\displaystyle\int_{} \ \epsilon \rho \frac{\ub^{n+1} - \ub^{n}}{\Delta t} \cdot \hat{\b{u}}\ \d \b{x} = \displaystyle\int_{} \bigg( \nablab \cdot \sigmab^{n+1}  + \epsilon \rho \boldsymbol{g}  - \displaystyle\sum_i \frac{H_i \b{F}_i^*}{V_i} \bigg)    \cdot \hat{\b{u}}\ \d \b{x}, \\
        &\displaystyle\int_{} \ \bigg(\nablab \cdot \big(\epsilon \ub^{n+1} + (1 - \epsilon) \b{v}^*\big)\bigg) \hat{p}\ \d \b{x} = 0,
    \end{align}

    As the mesh can reach fine resolutions to capture topology changes, the void fraction can tend to zero if an element is completely covered by a grain.
    In this case, the drag force tends to infinity, leading to a non-physical force.
    Additionally, the fluid acceleration, stress tensor and gravitational force terms vanish in the momentum equation.
    The only term remaining is the drag force, which imposes the velocity continuity in the grains domain.
    The pressure field is computed based on the velocity prediction in the continuity equation.
    As the prediction is based on the pressure gradient and the fluid velocity, the system of equations stays full-rank even though the void fraction tends to zero.
    In practice, to avoid large non-physical forces, a minimal value of $10^{-8}$ is set for the void fraction. 
\subsection{Advection}\label{advection}

\begin{figure}[H]
    \begin{center}
        \includegraphics[width=.7\textwidth]{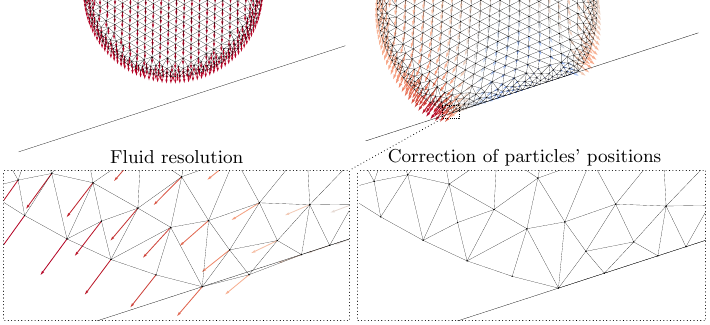}
        \caption{The position of the particles that cross a wall during the advection step is corrected by projecting them on the wall.}\label{fig:reproj}
    \end{center}
\end{figure}

The final step in the algorithm consists in displacing the fluid particles and solid grains using their obtained Lagrangian velocities. 
For the fluid, the positions are updated using a second-order scheme that approximates the acceleration of the particles using the difference in velocities between the previous and current time step. 
We consider the scheme proposed in \cite{suchde}. 
This choice for a second-order scheme is important in order to ensure better conservation properties near the interfaces. 
The grains are updated using a first-order implicit Euler method, which results from the iterative contacts solver.
\begin{align}
    \boldsymbol{x}^{n+1} &= \boldsymbol{x}^n + \boldsymbol{u}^{n+1} \Delta t  + \frac{\boldsymbol{u}^{n+1}-\boldsymbol{u}^{n}}{\Delta t} \Delta t^2\label{eq:fluid_displ}\\
    \boldsymbol{X}_{i}^{n+1} &= \boldsymbol{X}_{i}^n +  \boldsymbol{U}_{i}^{n+1} \Delta t, \label{eq:grains_displ}
\end{align}
For the fluid phase, a final step is required here to account for solid boundaries. 
Indeed, since the fluid boundaries are defined at the beginning of each time step, it is essential to ensure that boundary conditions are accurately applied despite the explicit nature of the position update.
In many PFEM approaches \cite{cerquaglia, onate} for fluid simulations, solid boundary particles are fixed on the walls, and define solid walls to which boundary conditions are then applied. 
The approach followed here, however, allows for a straightforward application of slip-boundary conditions along solid walls by considering only fluid particles and representing the walls only geometrically.
For particles which are, at the beginning of the time step, already on a wall, slip boundary conditions can be applied without any difficulty through the finite element formulations.
For fluid particles that cross a solid wall during a time step due to the explicit update, Equation \eqref{eq:fluid_displ}, a correction is made on these particles' velocities to project them on the boundary, as illustrated in Figure \ref{fig:reproj}.

\section{Validations}
In this section, two experiments serve to verify and validate the proposed method.
The first is a collapse of a partially filled water column, commonly referred to as a dam break. 
The second is a collapse of a column of grains into a water bed. 
The results are compared with references from the literature.  
In all of the experiments presented, the flow is driven by gravitational effects. 
Hence,  the Froude number is chosen as a dimensionless number to characterize the experiment as it measures the ratio of inertial forces to gravitational forces, 
$$ \Fr = \frac{v}{\sqrt{g h}}.$$
$v$ is a characteristic velocity, computed as an equivalent velocity if the water or granular column were to accelerate in free fall over a distance corresponding to half of its height $h$. 
The characteristic length and velocity that define the Froude number can then also be used to define non-dimensional times and lengths.

All the simulations presented next are simulated with free-slip boundaries along solid walls
\begin{equation*}
    \left\lbrace{\begin{array}{l}
        \ub \cdot \boldsymbol{n} = \boldsymbol{0}, \\
        \tb - (\tb\cdot\boldsymbol{n})\boldsymbol{n} = \boldsymbol{0},
    \end{array}}\right. 
\end{equation*}
where $\boldsymbol{n}$ is the outward facing normal and $\tb = \sigmab \cdot \boldsymbol{n}$ is the Cauchy stress vector.
In other words, a zero-velocity condition is imposed in the normal direction, and no stresses are applied in the tangential direction. A free surface is considered at the interface between water and air: 
\begin{equation*}
        \tb = p_0 \boldsymbol{n} - \kappa \gamma \boldsymbol{n},
\end{equation*}
where $p_0$ is an external atmospheric pressure, $\kappa$ is the curvature at the free surface, and $\gamma$ is the surface tension coefficient.

\subsection{Dam break}

\begin{figure}[H]
    \centering
    \includegraphics[width=\textwidth]{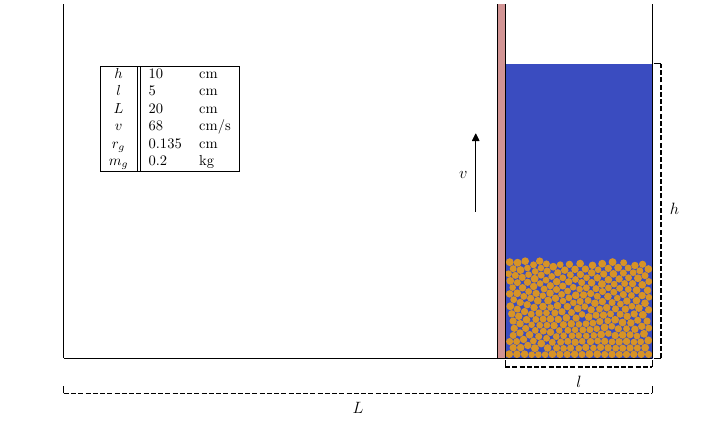}
    \caption{Dam break test case setup.}\label{fig:dambreak_setup}
\end{figure}

In this first verification test case, a water column is initially at rest between three walls.
A set of grains are immersed into the water column.
The total mass of the grains is $200$g, and their radius is $1.35$mm, resulting in a total of 279 grains.
A Coulomb friction law is used to model tangential interactions between grains with a friction coefficient of $0.2$.
No friction is considered between the grains and the walls.
Surface tension is considered at the water-air interface, with a coefficient of $0.0072$N/m.
The wall left to the water column is a vertically sliding gate, which is lifted at $t=0$ with a velocity of $0.68$ m/s. 
The water column and immersed grains are then free to slide under the gate, into the empty portion of the container.
The initial setup with dimensions of the water column is presented in Figure \ref{fig:dambreak_setup}.
Along the walls, a free-slip boundary condition is imposed.

\begin{figure}[H]
    \centering
    \includegraphics[width=0.8\textwidth]{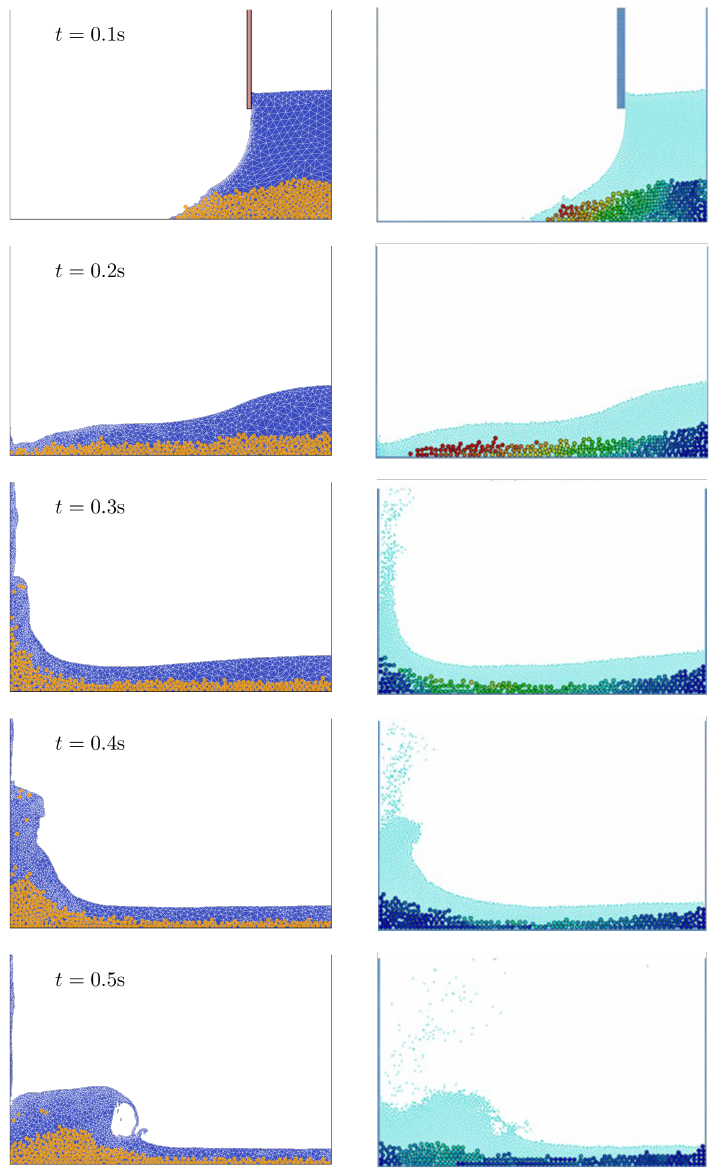}
    \caption{Time shots of the dam break simulation.
    On the left, results from the PFEM-DEM method are shown, and on the right, results from the SPH-DEM method are shown from Xie \cite{xie}.}\label{fig:dambreak_snapshots}
\end{figure}

The experiment can be divided into three main steps. 
First, the water and grain mixture slides into the open space. 
Two fronts, the fluid and the grain front, travel along the bottom in a clearly distinguishable wave. 
The timing of this first phase is from around 0 to 0.2s. 
Then, the wave hits the left wall, causing the mixture to rise and eventually collapse onto itself.
This generates high levels of turbulence, leading to chaotic phenomena such as splashing and trapped air bubbles.
The final step is a chaotic sloshing motion in the container, which eventually reduces to reach a final, steady state.
Figure \ref{fig:dambreak_snapshots} shows snapshots of the flow at different times and compares them to the results from \cite{xie}.
Movie 2 in supplementary materials shows the full simulation of the dam break.
In their work, a SPH-DEM formulation is used to simulate the same problem. 
The results are in good agreement, showing the potential of the PFEM-DEM method to capture such complex flows with the flexibility to keep a coarser mesh close to the grains.
However, a fine mesh is still required to capture the free surface dynamics accurately.
As the mesh nodes are advected, the volume of the fluid is not conserved.
In this benchmark, the fluid volume is reduced up to 2.5\% of its initial volume.

\begin{figure}[H]
    \centering
    \includegraphics[width=\textwidth]{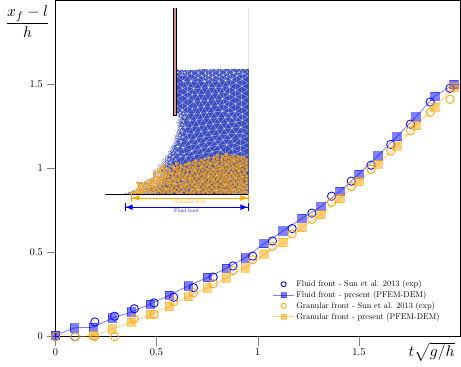}
    \caption{Time evolution of the granular front in orange and water front in blue for the dambreak.
    Numerical results, continuous lines, and squares, are compared to the experimental results, circles, from Sun \cite{sun}.}\label{fig:dambreak_fronts} 
\end{figure}

A quantitative study of this flow is feasible in the first step of the experiment, but becomes more delicate in the following ones as the flow takes a chaotic turn. 
To compare the results with the literature, the front positions of the fluid and grains are tracked.
The front is defined as the position of the particle with the lowest $y$ coordinate in the fluid or grain phase.
To characterize the flow, the dimensionless Froude number is used, leading to dimensionless time and space coordinates based on the initial height of the column and the gravitational acceleration.
Figure \ref{fig:dambreak_fronts} shows the evolution of the front positions with respect to time.
The accuracy of the PFEM-DEM method is demonstrated by the good agreement with the experimental results from \cite{sun}.

\subsection{Granular column collapse}
In the second benchmark, a granular column, initially dry, collapses into a water bed.
Similarly to the previous test case, a gate initially withholds the column, and at $t=0$, is lifted upwards with a velocity of 1 m/s.  
The grains then fall into the water bed, causing a free surface impulse wave to travel in the direction opposite to the grain column.\\ 
A physical experiment was carried out in \cite{sarlin} to study the influence of different size parameters of both the grain column and water bed on the granular front as well as the resulting free surface wave. 
This allowed them to qualify the type of grain collapse, and to predict the resulting free surface wave on the water.

\begin{figure}[H]
    \centering
    \includegraphics[width=\textwidth]{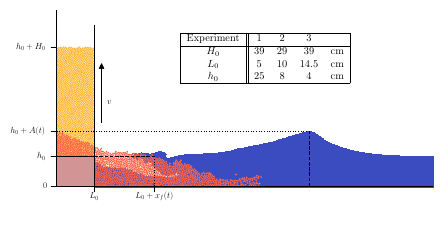}
    \caption{Granular column collapse test case setup.}\label{fig:landslide_setup}
\end{figure}

Three different experiments have been carried out, of which the main dimensional parameters are presented along with the initial setups in Figure \ref{fig:landslide_setup}.
For the granular phase, spherical grains with an average diameter of 5 mm are used, with a polydispersity coefficient of $1.2$. 
The density of the grains is $2500$ kg/m$^3$, and a Coulomb friction law with a coefficient equal to $0.9$ defines the friction between all rigid bodies.
For water phase, free-slip boundary conditions are considered along the solid walls, and the surface tension coefficient along the free surface is set to $0.0072$N/m.

An initial Froude number, as defined above, can be used to qualify the flow.
However, as suggested in \cite{robbe}, a better choice of characteristic velocity than the equivalent free-fall velocity, is the maximum horizontal velocity $v_{fm}$ of the granular front. 
\begin{align*}
    \Fr_f = \frac{v_{fm}}{\sqrt{g h_0}}     
\end{align*}
The reason for this different choice of velocity to describe the flow comes from the fact that the main driving force of the free surface wave is the horizontal push of the grains, and not their vertical drop as the columns collapses.
Though this velocity depends on the initial height of the column, it also depends on its initial width $L_0$. 
The work of \cite{sarlin} has allowed them to distinguish three different regimes as follows, based on the Froude number  ($\Fr_f$) defined with the maximum horizontal velocity of the front:

\begin{itemize}
    \item $\Fr_f \lesssim 0.35$: Nonlinear transition wave
    \item $0.35 \lesssim \Fr_f \lesssim 0.87$: Solitary non-breaking wave
    \item $0.87 \lesssim \Fr_f$: Transient bore wave
\end{itemize}

To validate the proposed method, three different experiments have been carried out, each corresponding to one of the three regimes.
The amplitude of the wave and the granular front position are tracked to validate the wave regime and the grains dynamics. 
The results are compared to the experimental results from \cite{sarlin}. 
The setup of each experiment, as well as the corresponding Froude numbers, are summarized in the table of Figure \ref{fig:landslide_setup}.

In the first case, the first regime is indeed observed as the free surface is dominated by a bore wave, as seen in Figure \ref{fig:landslide:snapshots1}. 
The full simulation video can be found in Movie 3 of supplementary materials.
In this case, grains become fully submerged in the fluid and the granular front goes to zero.
In the second case, the solitary non-breaking wave is observed generated by the granular motion, as seen in Figure \ref{fig:landslide:snapshots2}, with Movie 4 of the supplementary materials showing the full simulation. The fall of grains is not sufficient to submerge all the grains and the granular front is still visible once the wave has passed.
The wave takes time to roll over itself, and the wave amplitude remains close to a constant value until then. 
In the last case, the nonlinear transition wave is observed as seen in Figure \ref{fig:landslide:snapshots3} and Movie 5 of supplementary materials. 
The grains motion expelled a significant amount of fluid, leading to a nonlinear wave regime which quickly rolls over itself and breaks.
A part of the fallen grains remains dry as the water has been expelled and two internal friction angles can be observed in the grains, one for the dry grains, and one for the submerged grains.

Qualitatively, the three regimes are accurately captured by the numerical simulations.
However, disparities are observed in the granular front position, but this can be attributed to the uncertainty in the measure of the granular front position.
Indeed, to measure the granular front, the furthest grain at a height of $y=h_0$ is taken, a illustrated in Figure \ref{fig:landslide_setup}.
With the aim of computing the maximum horizontal velocity of the grains at the free-surface, responsible of generating the impulse wave, this measure is the most relevant option.
However, this choice does not lead to consistent results if grains are fully submerged in the fluid, as the granular front is not visible anymore.
If one is only interested in computing the maximum position reached by the grains inside the flow, an alternative, more robust measure would be to take the runout distance of the avalanche, defined as the position of the farthest grain still connected to the contact network \cite{coppin2023collapse}.
The order of magnitude of the wave amplitude is well captured, but the amplitude is slightly overestimated.
This can be attributed to a higher grains velocity in the numerical simulations, as the grains are not subject to the same friction as in the physical experiments.
Although the three regimes are correctly captured qualitatively and respect the bounds, we note some discrepancies with the obtained Froude numbers. 
Different reasons can be enumerated to explain these differences: 
three-dimensional effects, the parametrization of the grains' friction, the measurement method for computing the granular front position and velocity.
From the numerical setups used in this paper, the obtained values for $\Fr_f$ are presented in table \ref{tab:froude}.
Although the results are not exactly matching those of the experiments, the regimes are still detected correctly within the bounds proposed by \cite{sarlin}.

\begin{table}[H]
    \begin{center}
        \begin{tabular}{ |c||c|c| } 
            \hline
            Experiment  & $\Fr_f$ (PFEM-DEM) & $\Fr_f$ (exp., Sarlin et al.) \\
            $1$ & 0.27 & 0.19\\
            $2$ & 0.83 & 0.65\\
            $3$ & 1.93 & 1.39\\
            \hline
        \end{tabular}
        \caption{Froude numbers based on the maximum horizontal velocity of the grain column for the different cases, compared with the experimental data.}\label{tab:landslide_regimes}\label{tab:froude}
    \end{center}
\end{table}

\begin{figure}[H]
    \centering
    \includegraphics[width=\textwidth]{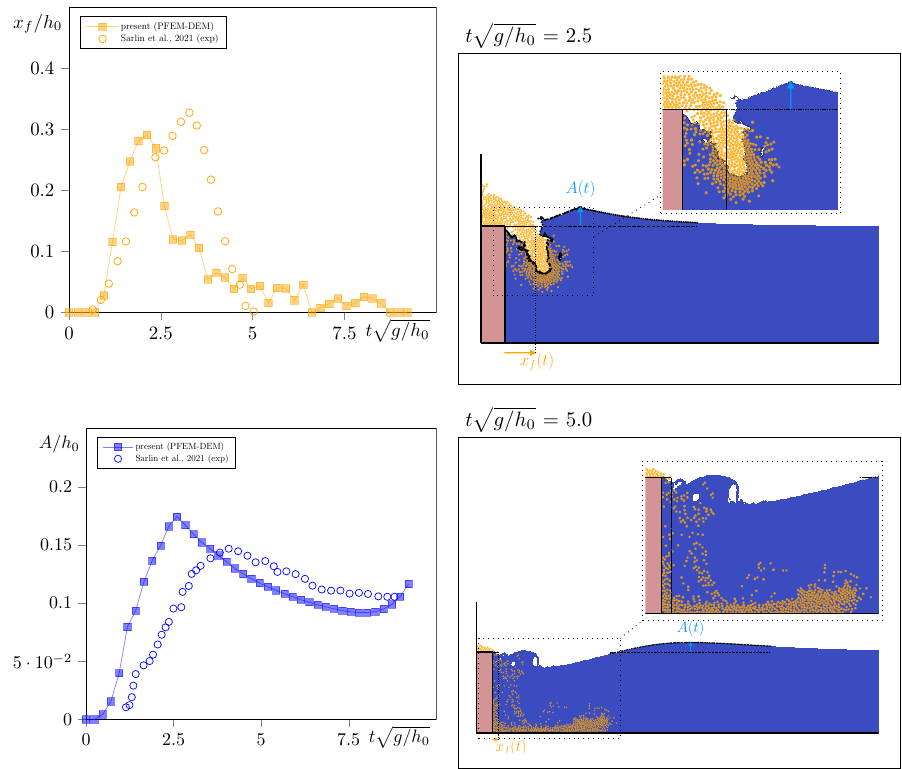}
    \caption{Granular collapse: case 1}\label{fig:landslide:snapshots1}
\end{figure}

\begin{figure}[H]
    \centering
    \includegraphics[width=\textwidth]{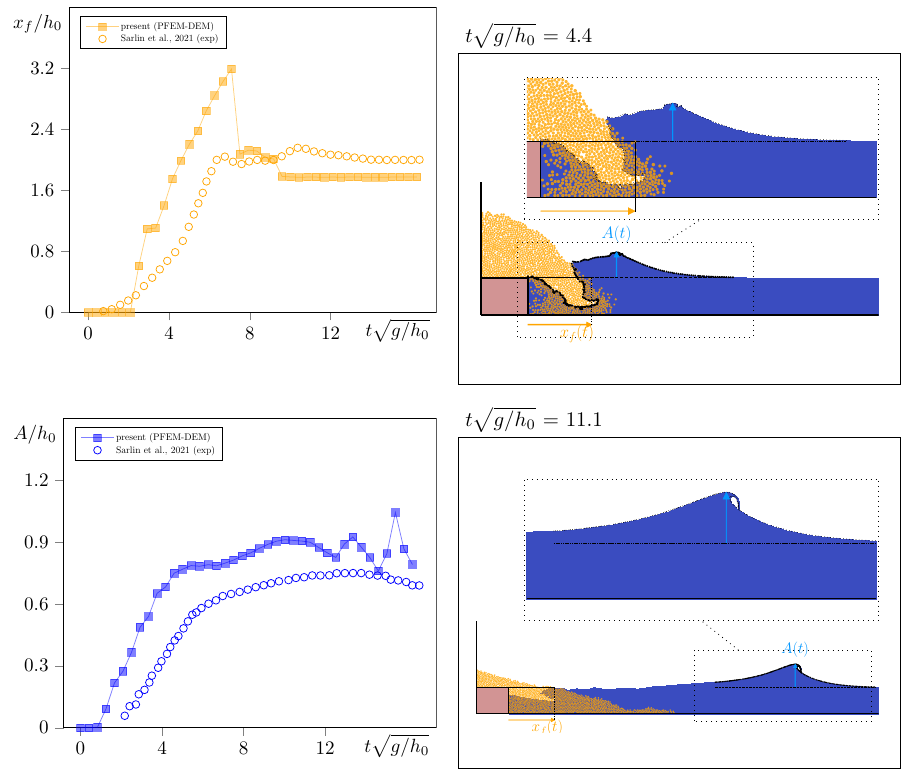}
    \caption{Granular collapse: case 2}\label{fig:landslide:snapshots2}
\end{figure}

\begin{figure}[H]
    \centering
    \includegraphics[width=\textwidth]{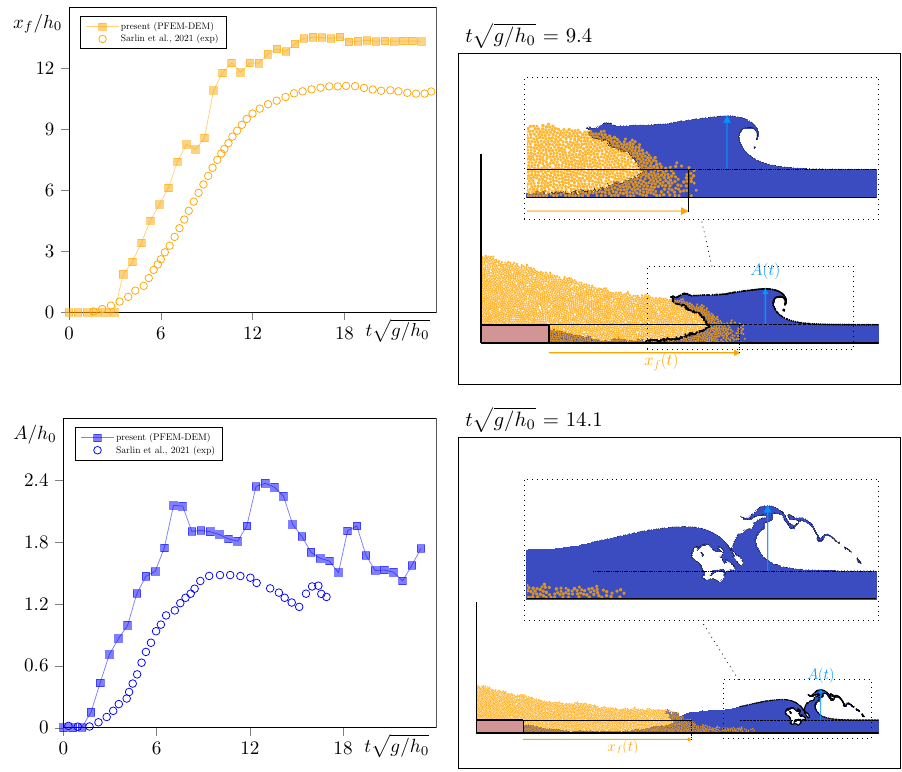}
    \caption{Granular collapse: case 3}\label{fig:landslide:snapshots3}
\end{figure} 

\section{Lituya bay landslide and tsunami}

To investigate the robustness and accuracy of the proposed model, a realistic case, the Lituya Bay landslide and tsunami, is studied.
On 10 July 1958, an earthquake along the Fairweather fault caused a large landslide above Lituya Bay in Alaska.
The loose rock slid downhill and hit the waters of Lituya Bay, creating a $60$m wave that reached heights of over $500$m on the opposite shore. 
The tsunami resulted in human casualties and homes were destroyed.
As mountains become increasingly unstable due to climate change, such events are expected to become more frequent.
Understanding and predicting the impact of such events is crucial to mitigate the risks.
The presented numerical method is a suitable tool for such phenomena, as both the landslide and the tsunami can be simulated and analyzed, providing a fine description of both the grain dynamics and the wave motion.
To quantitatively compare the results with the real event, the numerical results are compared with the experimental results of \cite{fritz}.
In their study, a 1:675 scale experiment of the Lituya Bay landslide and tsunami was carried out in order to better quantify the event.

\begin{figure}[H]
    \centering
    \includegraphics[width = \textwidth]{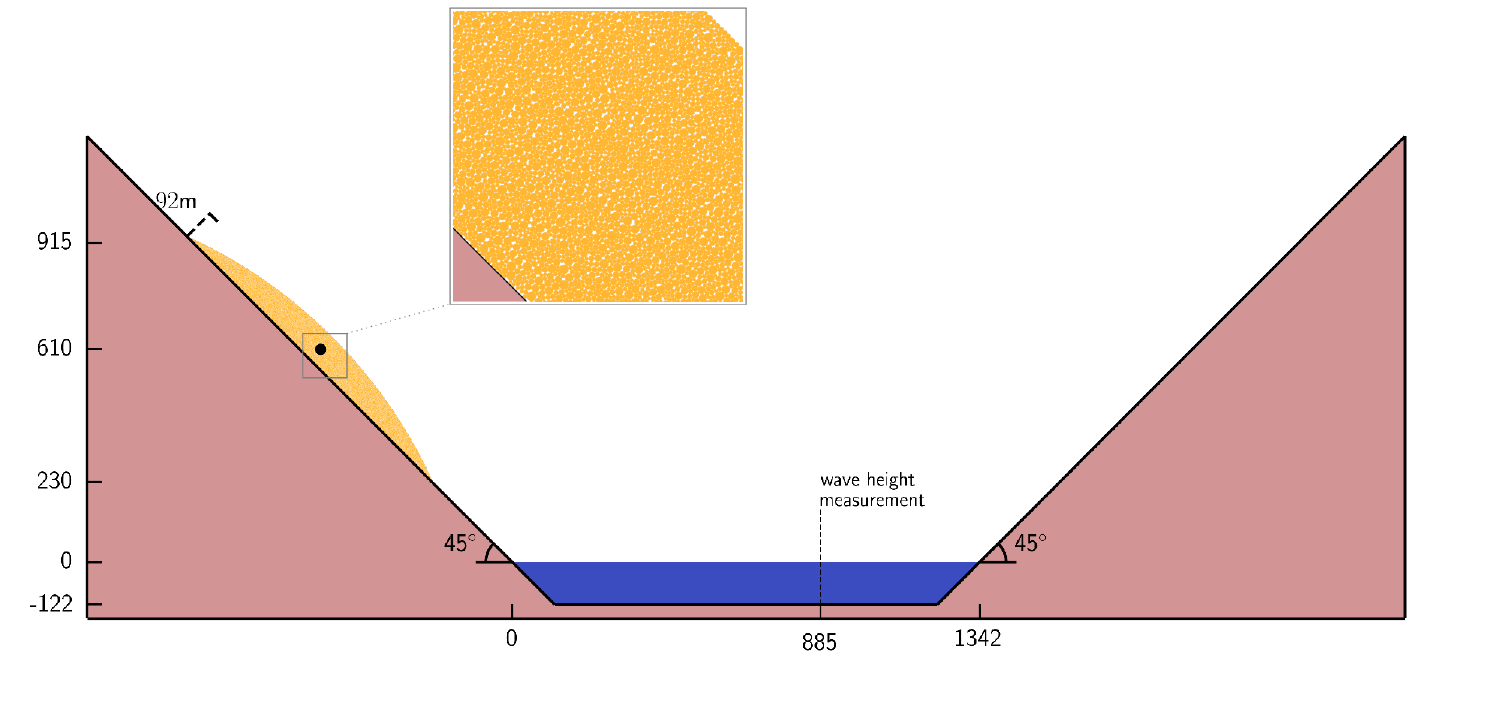}
    \caption{Lituya bay landslide and tsunami setup.}\label{fig:lituya_setup}
\end{figure}

The setup of the experiment is shown in Figure \ref{fig:lituya_setup}.
According to the estimated amount of rock that was torn from the hill, a mass of grains 970m long and 92m thick, is positioned 230m above the water level. 
This gives a total mass per unit thickness of $98.5 \times 10^3$ t/m.
The grains have a density of 2.64 t/$\text{m}^3$, and a polydispersity ratio of 3 allows the bulk to be more compact.
The average grains radius is set to 1.35m, in order to match the 2 mm radius used in \cite{fritz}, put to real-life scale.
This results in a total of 27960 grains.
The friction coefficient between the grains and with the walls is set to $\mu = 0.93$, which corresponds to an effective internal friction angle of 43$^\circ$.  
The water bed, representing a cross section of the Gilbert Inlet in the Lituya Bay, has a depth of 122 m, and a total length of 1342 m. 
Both shores rise with an angle of 45$^\circ$.
At $t=0$, the grain mass is set in motion. 
The Froude number is here also the relevant dimensionless number to relate the impact of the grains to the potential energy stored in the water bed. 
Considering the mean velocity $v_g$ of the grains as they impact the water, estimated at $110$m/s in \cite{fritz}, and the water depth of the bay, equal to $122$m, the resulting Froude number is equal to
$$ \text{Fr} = \frac{v_g}{\sqrt{gh}} = 3.18.$$

In this case, the wave is expected to break, as the Froude number is largely superior to the limit proposed in the previous testcase for the transition from non-breaking to breaking waves. 
The numerical experiment has been calibrated in order to match this Froude number.

\begin{figure}[H]
    \centering
    \includegraphics[width=\textwidth]{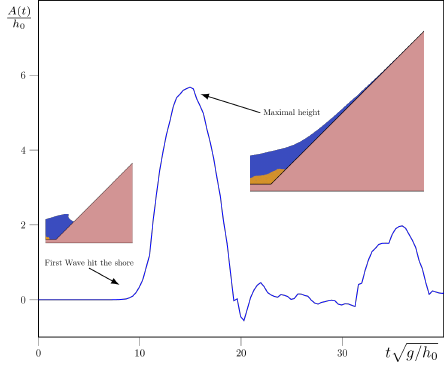}
    \caption{Evolution of the wave amplitude on the opposite shore. Dimensionless height is defined by the initial depth $h_0$ and the gravitational accelerating is used to define a dimensionless time.}\label{fig:lituya_wave_shore}
\end{figure}

\begin{figure}[H]
    \centering
    \includegraphics[width =\textwidth]{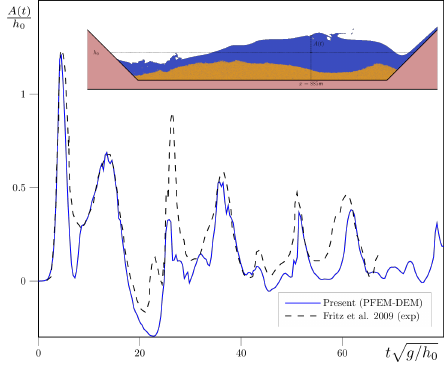}
    \caption{Wave height evolution at $x=885$m. Numerical results, the continuous blue line, is compared to the experimental results from \cite{fritz}, dotted black curve.
    The dimensionless height and time are computed using the initial bathymetry $h_0$ as the reference height and $g$ the gravitational acceleration.}\label{fig:lituya_wave_height}
\end{figure}

Some snapshots of the simulation are presented in Figure \ref{fig:lituya_snapshots}.
Movie 6 in the supplementary materials shows the full simulation.
The wave hits the opposite shore before breaking, as has also been observed in \cite{fritz}.
The height reached on the opposite shore is tracked in Figure \ref{fig:lituya_wave_shore}.
Observations at Lituya Bay stated the inertia of the wave was sufficient to damage trees at over $500$m.
The numerical model indeed shows that such a height is reached by the wave.
The maximal point reached a height of around $750$m. 
This value may be overestimated as water is assumed to slide freely along the boundary.
When the wave returns in the opposite direction, a sloshing phenomenon appears, and the wave reflects multiple times against the opposing shores.
To track if the model is able to capture these reflectives waves, the wave amplitude at a position of $885$m from the initial shore is compared to the experimental data from \cite{fritz} at Figure \ref{fig:lituya_wave_height}.
The dimensionless height and time are computed using the initial water depth $h_0$ as the reference height and $g$ the gravitational acceleration.
The Lagrangian formulation is able to accurately capture the main trends in the wave dynamics.
However, as the domain is advected and meshed at each iteration, the water volume is not conserved. 
The initial impact of the grain on the water bed leads to high deformations and the formation of a first wave.
These deformations are the main source of volume loss. 
The maximal volume variation is around 10$\%$ of the initial volume.

\begin{figure}[H]
    \centering
    \includegraphics[width = \textwidth]{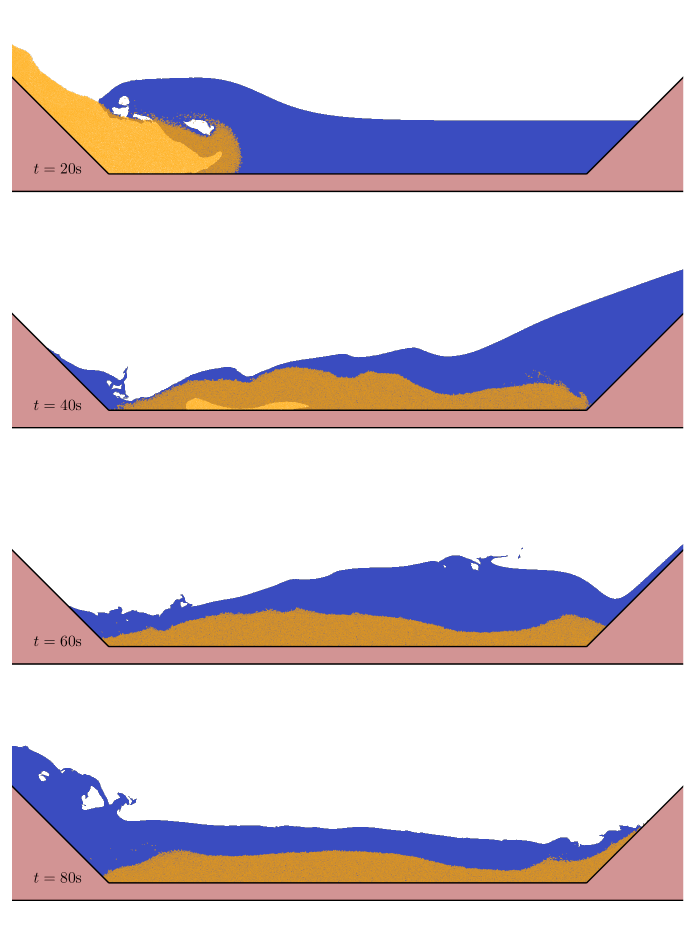}
    \caption{Snapshots of the Lituya bay landslide and the wave generated by the fallen rocks at time $20$, $40$, $60$ and $80$s.}\label{fig:lituya_snapshots}
\end{figure}

\section{Conclusions}

This paper presents a coupled Lagrangian model to simulate immersed granular flows with free surfaces and moving domains.
The method is based on a coupling between a particle finite element method (PFEM) solver and a discrete element method (DEM) solver.
The coupling provides robustness and flexibility to the simulation.
As the PFEM solver is based on a Lagrangian formulation, it can handle strongly deforming domains and free surfaces.
The DEM solver, on the other hand, is well suited for granular problems and provides insight into the contact dynamics.
The Volume-Averaged formulation avoids the need to provide constitutive laws for the granular phase as it is directly captured by the contact dynamics.
However, a constitutive law is required to model the fluid-grain interaction.
In addition, the mesh size can be chosen independently for each phase, allowing the mesh to be refined only where it is needed.
In practice, a fine mesh is used close to the free surface to capture large deformations and topological changes, while a coarser mesh is used in the bulk of the domain.

The accuracy of the method is assessed on a few benchmarks.
First a dam break consisting of grains and water is simulated to serve as a verification of the numerical model.
The results show good agreement with experimental data and existing numerical methods to track the free surface and the granular front.
Next, to investigate the behaviour of the method in the presence of strong deformations, a granular column collapse is simulated.
Depending on the geometry of the granular column and the depth of the water bed, leading to different fall velocities of the grains, the wave dynamics can be very different. 
Experimental results from the literature \cite{sarlin} proposed a characterization based on the flow's Froude number, and divided the experiments into three regimes.
These three regimes are well captured by the proposed method, validating the method and showing the capability of the approach to solve complex phenomena such as high amplitude free-surface waves triggered by landslides.

Finally, the method is applied to a realistic event, the 1958 Lituya Bay landslide and tsunami.
The results show that the method can capture the main features of the event, such as the landslide and the subsequent tsunami.
Moreover, the complex behaviour is still well captured when the wave begins to reflect on the opposite shore.

The results presented in this work show that the method can be applied to a wide range of problems, but there are still some limitations and room for improvement.
For one thing, the fluid's volume is not conserved.
This error is mainly due to the fact that the node advection scheme used in the method is not conservative.
Additionally, remeshing at each time step introduces some numerical diffusivity, and the fact of defining the domain based on the $\alpha$-shape of the point cloud can cause unphysical deformations.
This can lead to some inaccuracies in the results, especially if the free surface is not fully resolved.
To mitigate this, the mesh size is refined close to the free surface.
At present, the method is limited to two dimensions, but an extension to three dimensions is planned and no major differences are expected.
The multi-scale approach can also have some limitations as its computational cost scales with both the number of grains and the number of fluid particles.
This can become a bottleneck for real scale applications as the number of particles can become very large.
In this case, a parallel implementation of the method is required, or the use of a continuous representation of the grains.

Although the applications are limited to landslide-related phenomena in this paper, the approach can be exploited on a wide variety of physical experiments. 
For example, it could be applied to mixing problems in tanks, soil digging problems, or even small-scale applications such as injector-induced bubble flows. 

\section{Acknowledgments}
The authors acknowledge that Michel Henry and Thomas Leyssens contributed equally both to the research and writing of this paper. 
Thomas Leyssens is a Research Fellow of the F.R.S.-FNRS.
Computational resources have been provided by the supercomputing facilities of the Université catholique de Louvain (CISM/UCL) and the Consortium des Équipements de Calcul Intensif en Fédération Wallonie Bruxelles (CÉCI) funded by the Fond de la Recherche Scientifique de Belgique (F.R.S.-FNRS) under convention 2.5020.11 and by the Walloon Region.

\printbibliography
\end{document}